\documentclass[twocolumn,epsfig,showpacs,preprintnumbers,amsmath,amssymb]{revtex4}
\makeatletter
\def\@dotsep{4.5}
\makeatother
\usepackage[dvips]{graphicx}  
\usepackage{amsmath}
\usepackage{color}
\definecolor{textcolor}{cmyk}{0,0,0,1}
\definecolor{magenta}{rgb}{1,0,1}
\definecolor{green}{rgb}{0,1,0}
\definecolor{red}{rgb}{1,0,0}

\begin{document}
\draft
\title{Anomalous exchange interaction between intrinsic spins in conducting graphene systems}

\author{H. Santos$^{1}$}
\email{hernan.santos@fisfun.uned.es}
\author{David Soriano$^2$}
\author{J.J. Palacios$^{3}$}
 
\affiliation{
$^1$ Departamento de F\'{\i}sica Fundamental, Universidad Nacional de Educaci\'on a Distancia, Apartado 60141, E-28040 Madrid, Spain\\
$^2$ ICN2 - Institut Catal{\`a} de Nanoci{\`e}ncia i Nanotecnologia, Campus UAB, 08193 Bellaterra (Barcelona), Spain\\
$^3$ Departamento de F\' isica de la Materia Condensada, Condensed Matter Physics Center (IFIMAC),
and Instituto Nicol\'as Cabrera (INC), Universidad Aut\'onoma de Madrid, Madrid 28049, Spain.}

\date{\today}

\begin{abstract}
We address the nature and possible observable consequences of singular one-electron states that appear when strong defects are introduced in the metallic family of graphene, namely, metallic carbon nanotubes and nanotori. In its simplest form,  after creating \textit{two} defects on the same sublattice, a state may emerge at the Fermi energy presenting very unusual properties: It is unique, normalizable, and features a wave function equally distributed around both defects.  As a result, the exchange coupling between the magnetic moments generated by the two defects is anomalous. The intrinsic spins couple ferromagnetically, as expected, but do not present an antiferromagnetic excited state at any distance. We propose the use of metallic carbon nanotubes as a novel electronic device based on this anomalous coupling between spins which can be useful for the robust transmission of magnetic information at large distances.
\end{abstract}
\pacs{73.22.-f, 73.73.-b, 75.75.-c}

\maketitle

\emph{Introduction.} After almost one decade of research in graphene and graphene-based structures\cite{Castro-Neto} and more than two decades of research in carbon nanotubes (CNT's)\cite{Iijima_1991}, little remains to be known on the nature of the single-particle electronic states of these systems. Nowadays focus is shifting towards defect-dependent electronic properties. For large-scale practical applications understanding the role of defects is essential and much work has been done on the tuning of graphene properties through functionalization\cite{Elias,Lopez-Bezanilla,nn200558d} or the controlled manipulation of defects\cite{Rodriquez-Manzo,Lehmann,Meyer,Gomez-Navarro,Warmer}.

Vacancies and adsorbates play a central role in this game. Both types of defects act as a strong perturbation since both can give rise to localized states at or near the Dirac point. The basics of the emergence of these states lies in the bipartite nature of the graphene lattice and simple rules\cite{PhysRevB.49.3190}. For instance,  when one $p_z$ orbital is removed from the system in an otherwise perfect lattice, a zero-energy state must appear on the other sublattice, typically around the perturbation\cite{Liang}. Whether or not this is accompanied by the emergence of magnetism is a matter of current debate for vacancies since the passivation and  structural details become relevant\cite{Palacios12}. In the case of H adatoms, the situation is, however, much more clear\cite{Duplock,Soriano}. Keeping this in mind, we will generically refer to both unreconstructed vacancies and H adatoms simply as defects from now.  
\begin{figure}[b]
\includegraphics[width=0.36\textwidth,angle=-90,clip]{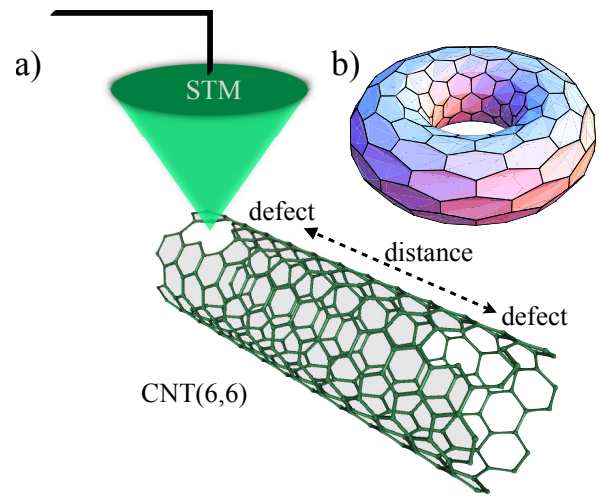}
\caption{ (Color online) Schematic views of a metallic nanotorus (upper figure) and of our proposed experimental set-up based on a metallic nanotube with two defects and a scanning tunneling microscopy tip.}
\label{setup}
\end{figure}

For infinite or gapful graphene (e.g., armchair-terminated flakes) the emergence of defect-induced ``zero-energy'' (or Fermi energy) states is well understood\cite{vozmediano:155121,PhysRevB.49.3190}. However, the question that we want to address here is: What happens when, previously to the introduction of defects, the bulk density of states (DOS) is already finite at zero energy? This situation naturally occurs in metallic CNT's and nanotori with appropriate radii (metallic nanoribbons unavoidably present a small gap\cite{son:216803} and are excluded from this discussion). We will show that while the presence of a first defect is not revealed in the DOS,  a second one on the same sublattice may induce a peak at the Fermi energy which corresponds to a state localized \textit{around both defects} regardless of their relative distance. Thereof the term \textit{bi-local} which will be used  to refer to such state from now on. The conditions for the appearance of this state in metallic CNT's and nanotori (as the ones shown in Fig. \ref{setup}) are analysed with a simple tight-binding model. The intrinsic spin induced by the defects and the anomalous behaviour of its magnetic coupling   is exposed through density functional theory (DFT) calculations. A possible use of such a peculiar electronic state for transmission of (magnetic) information without losses at long distances is proposed. Fig. 1(a) illustrates his possibility. There a magnetic probe (e.g., a scanning tunnelling microscope (STM) magnetic tip) near a defect on a metallic CNT is used to register changes in the local magnetic environment at  the second defect or to manipulate the magnetic state of this, but non-locally.

\begin{figure}
\includegraphics[width=0.58\textwidth,angle=-90,clip] {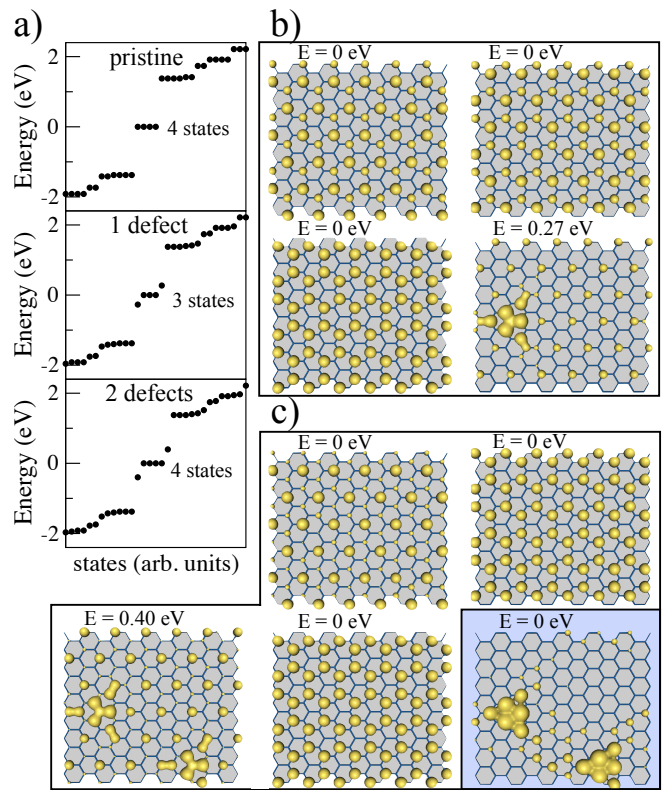}
\caption{(color online)
(a) Electronic spectrum close to zero energy in a CNT(6,6) nanotorus with $9$ unit cells.  (b) Wavefunctions for one defect. (c) The same for two defects that generate a bilocal state. The shaded panel shows this state.
}
\label{nanotorus}
\end{figure}

\textit{Bi-local states in carbon nanotori.-} Although their production in the lab is rare, we start our discussion with the help of a graphene nanotorus.  A nanotorus can be seen as a finite-length CNT with the two ends joined as to form a ring-like structure [see Fig. \ref{setup}(b)].  We first model the Hamiltonian of the $\pi$ carriers by a single first-neighbour hopping parameter $t$:
\begin{equation}
H = - t \sum_{i,j,\sigma}  c^\dagger_{i,\sigma} c_{j,\sigma},
\label{eq:ham}
\end{equation}
where $ c^\dagger_{i,\sigma}$ ($ c_{j,\sigma}$) is the creation
(annihilation) operator at atom $i$ ($j$) of a $\pi$ electron. We assume a value for the hopping parameter
between near-neighbor orbitals of $t=2.66$ eV. Neglecting curvature effects, the electronic structure of these systems can actually be inferred from those of infinite graphene by simply selecting the states in the first Brillouin zone that are compatible with the periodic boundary conditions of the nanotorus \cite{Saito_book,Hamada_1992,DresselhausPW_1996}. Depending on the nanotorus, the zero-energy states at the Dirac points will be part of this selection or not. In particular, a metallic CNT$(n,n)$ with the ends connected with a periodicity along the tube axis  being proportional to $3a$ (where $a$ is the graphene lattice parameter) is a ``metallic'' nanotorus with zero-energy states. The tight-binding energy spectrum obtained for a nanotorus of this kind is shown in the upper panel of Fig. \ref{nanotorus}(a). An infinitesimal sublattice symmetry breaking that splits the four-fold degeneracy into a pair of two-fold degenerate states with an electronic density confined to the A or B sublattice has been added for convenience.

We also show the low-energy spectrum [Fig. \ref{nanotorus}(a), middle and lower panels] and associated wavefunctions  [Figs. \ref{nanotorus}(b) and (c)] for one and two defects on the nanotorus. When a first defect is created, e.g., on the A sublattice, one  naively expects a zero-energy state to be created on the B sublattice\cite{PhysRevB.49.3190}. This, however, hybridizes with one of the existing A-states, forming a bonding-antibonding pair away from zero energy [see the states at $E=\pm0.27$ eV in middle panel of Fig. \ref{nanotorus}(a)] and reducing the number of zero-energy states down to three.  

When a second defect is added, two possible scenarios appear, depending on their relative position. First, the pair of defects may be positioned on the same sublattice with a relative vector $\vec{R}$ satisfying the natural periodicity of mixed-valley wave functions at the Dirac point:
\begin{align}
|\Phi (\vec{R})|^2 & = \cos{(\vec{K}-\vec{K'}) \cdot \vec{R}}  \notag + C \\ 
 & \rightarrow(\vec{K}-\vec{K'}) \cdot \vec{R}  = 2 n \pi ,
\label{condition}
\end{align}

where $C$ is a constant (This condition was already identified in Ref. \onlinecite{Thygesen,Khalfoun} as responsible for unusual scattering properties in nanotubes). Figure \ref{nanotorus}(c) shows an example of this first scenario in which the second defect does not affect the three states of the nanotorus. Instead, \textit{it creates a new bi-local state at zero-energy}, with an electronic density distributed mainly around the two vacancies on the B sublattice (see shaded panel in Fig. \ref{nanotorus}(c)]). In the second scenario (not shown) the second defect withdraws one of the three zero-energy states, remaining only two of them. 

We must stress the unusual nature and uniqueness of the bi-local state. Note that linear combinations of two localized states that are spatially far apart from each other may also give rise to bi-local states, apparently similar to the one we describe here. The difference lies in that these come in \textit{bonding-antibonding pairs}, which become degenerate for large distances. Any perturbation can break the degeneracy and localize the electronic density on either one of the original local states. These pairs are also present in our spectrum [see the states at $E=\pm0.4$ eV in lower panel of Fig. \ref{nanotorus}(a)]. In addition, they are not fully localized [notice the finite weight in all atoms of the nanotorus shown in Fig. \ref{nanotorus}(c)]. On the contrary, the electron wavefunction of the bi-local state is fully localized and intrinsically split into two locations. Since this state is unique, perturbations are not expected to easily change this fact.

\begin{figure}
\includegraphics[width=0.6\textwidth,angle=-90,clip]{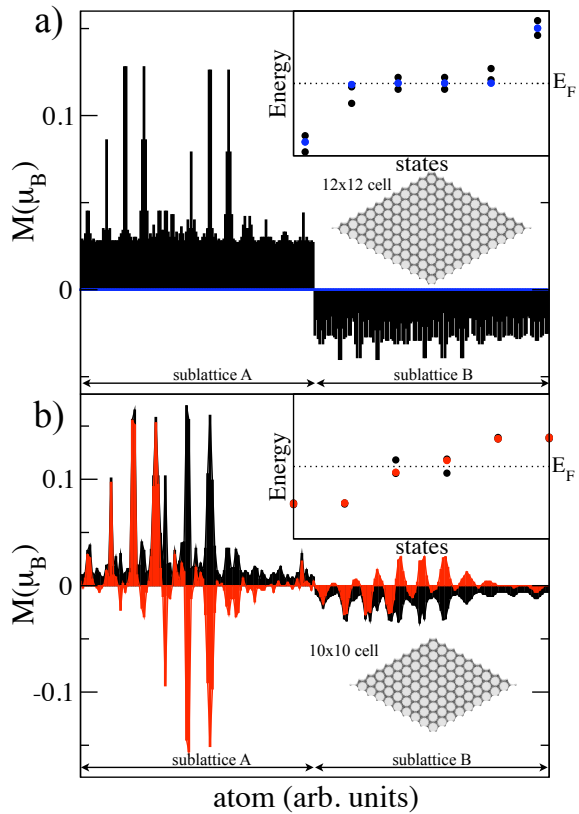}
\caption{(color online) Magnetic moments per atom for ferro (black) and antiferromagnetic (red) states of an AA pair in a metallic nanotorus (top) and semiconducting one (bottom). In the metallic case there is no antiferromagnetic state. The insets show the low-energy spectrum in the four cases, including the non-magnetic one (blue). }
\label{dft}
\end{figure}

It is known that localized states at the Fermi level spin-split under electron-electron interactions\cite{Louie,magnetic_bilayer} and the bi-local state is no exception. We have carried out spin-polarized DFT calculations with the SIESTA\cite{SIESTA} code for nanotori amenable to present bi-local states. To avoid irrelevant curvature effects which might interfere in the discussion we have actually performed standard calculations\cite{Datos} for a flat $3n\times 3n$ unit cell as the one shown in the inset of Fig. \ref{dft}(a), but only using the $\Gamma$ point. The defects are created by adsorption of H atoms. When the condition given by Eq. \ref{condition} is satisfied a spin density appears around both H atoms [see Fig. \ref{dft}(a)] and  closely following the density of the bi-local state shown by the tight-binding calculations. The integrated spin density amounts to $S=1$ as dictated by Lieb's theorem\citep{Lieb}. The remarkable fact here is that an antiferromagnetic state cannot be created by reversing the spin orientation around one H atom since the spin density is essentially supported by the bi-local state which is unique. One could say that the exchange coupling between the local magnetic moments around the H atoms is infinite since any attempt at generating an antiferromagnetic solutions ends up with a non-magnetic state of much higher energy (see blue line and dots in the inset). 
 This will likely require very large cells and we have not been able to verify this end even for our largest systems. All this contrasts to the finite exchange coupling that is always obtained in infinite graphene\cite{Soriano} or in gapful nanotori as shown in Fig. \ref{dft}(b). Finally notice that if the pairs of H atoms do not satisfy Eq. \ref{condition} itinerant or extended ferromagnetism is obtained for AA pairs and a non-magnetic state for AB pairs.

\textit{Bi-local states in CNT's.-} We now examine whether or not the bi-local states and the anomalous exchange couplings also appear in a much more common metallic CNT$(n,n)$. They are in many regards similar to the previously discussed nanotori, but with infinite circumference radius and open boundary conditions.  As before we are interested in the electronic  states that emerge after introducing two defects; therefore the system lacks translational invariance. We use a standard Green's
function (GF) approach to calculate the electronic structure\cite{Chico3,Jacob11,Munoz-Rojas06}, more specifically the DOS and the conductance. In regards to this, we split the system into three parts, namely a central region containing the defects and connected to the right and left leads. The Hamiltonian can thus be written as
\begin{equation}
H= H_C  + H_R + H_L + h_{LC} + h_{LR},
\end{equation}
where $H_C$, $H_L$, and $H_R$ are the Hamiltonians of the central
portion, left and right leads respectively, and $h_{LC}$, $h_{RC}$
are the hopping matrices from the left $L$ and right $R$ lead to the central
region $C$. The GF of the latter is
\begin{equation}
\mathcal{G}_C(E) = (E-H_C - \Sigma_L -\Sigma_R)^{-1},
\end{equation}
where $\Sigma_\ell= h_{\ell C}g_\ell h_{\ell C}^\dagger$ is the
selfenergy due to lead $\ell=L,R$, and $g_\ell = (E -H_\ell)^{-1}$
is the GF of the semiinfinite lead $\ell$.
Complementary to the local DOS, the conductance can also be computed as thoroughly described in the literature\cite{Santos,Jacob11}. 

\begin{figure}
\includegraphics[width=0.55\textwidth,angle=-90,clip]{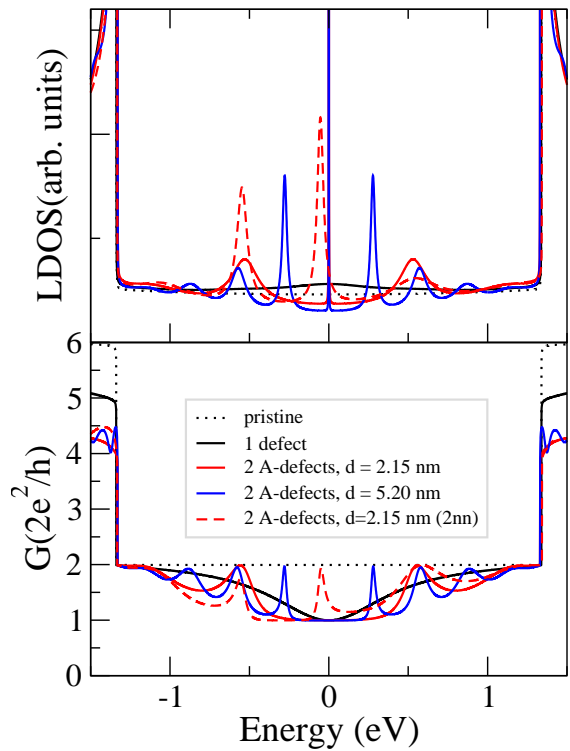}
\caption{(color online) DOS (top panel) and conductance (bottom panel) of a CNT(6,6) with different defects configurations as explained in the legend. The dashed lines correspond to second-near-neighbor hopping for one case.}
\label{nanotube}
\end{figure}

Figure \ref{nanotube} depicts the DOS and the conductance of a CNT(6,6) in the tight-binding approximation with different defects configurations.  A single defect positioned on the A (or B) sublattice barely shows up as a bump close to the Fermi energy ($E=0$ eV). This is analogous to the previous discussion for the nanotorus except for the fact that now we have a continuous DOS and it contrasts very much to the effect of the same defect on infinite graphene where a peak corresponding to a semi-localized state appears at the Dirac point\cite{Liang}. While the DOS does not reflect the defect, the conductance of the nanotube drops from 2$G_0$ to $G_0$ ($G_0=2e^2/h$)\cite{Chico3,Thygesen}. From the nanotorus results this can be understood as due to the complete blocking of one of the channels since current cannot be carried by states which only live on one sublattice. 

\begin{figure}
\includegraphics[width=0.28\textwidth,angle=-90,clip]{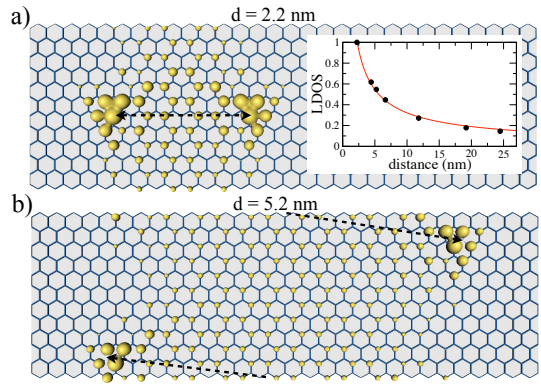}
\caption{DOS at zero energy projected on all atoms for a CNT(6,6) (unrolled view) with two defects positioned at a distance of 2.15 nm (a) and 5.2 nm (b) from each other. The inset shows the decay of the maximum LDOS around the defects as a function of the relative distance.}
\label{DOS}
\end{figure}

When the second defect is located on the same sublattice satisfying Eq. \ref{condition}, a zero-width peak now appears at zero energy. The associated state is fully localized and does not hybridize with the conduction electrons as anticipated from the nanotorus results. Notice also that, as Fig. \ref{nanotube} shows, changing the distance between defects does not have a significant effect. The zero-width peak remains located at zero energy, although the oscillations in the DOS grow and the peaks close to zero energy sharpen.  Also as expected, defect pairs which do not satisfy Eq. \ref{condition} do not induce a zero-energy peak. 
Figure \ref{DOS} depicts the local DOS at $E=0$ eV for the two cases.  A strongly localized density \textit{around both vacancies} appears. As expected for a zero-energy state, this wave function has zero weight in the sublattice hosting the defects. As far as we have been able to check, when the distance is increased the exotic bi-local character of this wavefunction persists with a slowly decreasing weight around defects [see inset in Fig. \ref{DOS}(a)]. 

One can foresee that, due to the finite DOS at the Fermi energy, a more realistic CNT Hamiltonian, which will always break electron-hole symmetry, may have an effect on the localization of the bi-local state. The simplest way to do this is by adding a second-nearest neighbour interaction ($t'=0.1 eV$) to our previous tight binding calculations. In Fig. \ref{nanotube} we show with dashed lines the results for the  DOS and the conductance for one of the cases. Notice that the bi-local state is still close to the Fermi energy (which is now at $E=0.585$ eV, but has been shifted to zero for clarity) acquires a finite width which indicates that it now couples to the continuum. Its nature, however, remains the same. This coupling may have important consequences since now this state can be electronically probed in transport, as shown by the conductance peak at that energy which restores the maximum conductance of the CNT(6,6). 

In the light of so many similarities, we do not expect significant differences in the behavior of the exchange couplings between nanotubes and nanotori. We have also verified through DFT calculations the impossibility of generating an antiferromagnetic state out of an AA pair of defects creating a bi-local state in the nanotube. All this encourages to propose the  experimental set-up shown in Fig. \ref{setup}. A STM magnetic tip located near a defect can be used to detect the appearance of a new defect at any distance from the tip or even local changes in the magnetic environment of this second defect. The tip could also manipulate the spin density orientation at the second defect from the distance. To our knowledge, this robust non-local transmission of (magnetic) information is unique to these systems, opening the possibility for a new class of carbon nanodevices where magnetic information can be non-locally stored. Further work should examine this proposal for a finite concentration of defects where multi-local instead of bi-local zero-energy states appear. 


\begin{acknowledgments}
This work was supported by MICINN under Grants Nos. FIS2010-21883 and CONSOLIDER CSD2007-0010, by Generalitat Valenciana under Grant PROMETEO/2012/011,  and by Spanish Ministry of Economy and Competitiveness (MINECO) under Grants No. FIS2010-21282-C02-02 and MAT2012-33911. H. S. gratefully acknowledge helpful discussions with Leonor Chico and Jose Enrique Alvarellos. J.J.P. and D. Soriano acknowledge discussions with M. A. Vozmediano in the initial stages of this work and also acknowledge computational support from the CCC of the Universidad Aut\'onoma de Madrid.
\end{acknowledgments}

\bibliography{bib_bilocal}

\begin{thebibliography}{32}
\expandafter\ifx\csname natexlab\endcsname\relax\def\natexlab#1{#1}\fi
\expandafter\ifx\csname bibnamefont\endcsname\relax
  \def\bibnamefont#1{#1}\fi
\expandafter\ifx\csname bibfnamefont\endcsname\relax
  \def\bibfnamefont#1{#1}\fi
\expandafter\ifx\csname citenamefont\endcsname\relax
  \def\citenamefont#1{#1}\fi
\expandafter\ifx\csname url\endcsname\relax
  \def\url#1{\texttt{#1}}\fi
\expandafter\ifx\csname urlprefix\endcsname\relax\def\urlprefix{URL }\fi
\providecommand{\bibinfo}[2]{#2}
\providecommand{\eprint}[2][]{\url{#2}}

\bibitem[{\citenamefont{Castro~Neto et~al.}(2009)\citenamefont{Castro~Neto,
  Guinea, Peres, Novoselov, and Geim}}]{Castro-Neto}
\bibinfo{author}{\bibfnamefont{A.~H.} \bibnamefont{Castro~Neto}},
  \bibinfo{author}{\bibfnamefont{F.}~\bibnamefont{Guinea}},
  \bibinfo{author}{\bibfnamefont{N.~M.~R.} \bibnamefont{Peres}},
  \bibinfo{author}{\bibfnamefont{K.~S.} \bibnamefont{Novoselov}},
  \bibnamefont{and} \bibinfo{author}{\bibfnamefont{A.~K.} \bibnamefont{Geim}},
  \bibinfo{journal}{Rev. Mod. Phys.} \textbf{\bibinfo{volume}{81}},
  \bibinfo{pages}{109} (\bibinfo{year}{2009}),
  \urlprefix\url{http://link.aps.org/doi/10.1103/RevModPhys.81.109}.

\bibitem[{\citenamefont{Iijima}(1991)}]{Iijima_1991}
\bibinfo{author}{\bibfnamefont{S.}~\bibnamefont{Iijima}},
  \bibinfo{journal}{Nature} \textbf{\bibinfo{volume}{354}}, \bibinfo{pages}{56}
  (\bibinfo{year}{1991}).

\bibitem[{\citenamefont{Elias et~al.}(2009)\citenamefont{Elias, Nair,
  Mohiuddin, Morozov, Blake, Halsall, Ferrari, Boukhvalov, Katsnelson, Geim
  et~al.}}]{Elias}
\bibinfo{author}{\bibfnamefont{D.~C.} \bibnamefont{Elias}},
  \bibinfo{author}{\bibfnamefont{R.~R.} \bibnamefont{Nair}},
  \bibinfo{author}{\bibfnamefont{T.~M.~G.} \bibnamefont{Mohiuddin}},
  \bibinfo{author}{\bibfnamefont{S.~V.} \bibnamefont{Morozov}},
  \bibinfo{author}{\bibfnamefont{P.}~\bibnamefont{Blake}},
  \bibinfo{author}{\bibfnamefont{M.~P.} \bibnamefont{Halsall}},
  \bibinfo{author}{\bibfnamefont{A.~C.} \bibnamefont{Ferrari}},
  \bibinfo{author}{\bibfnamefont{D.~W.} \bibnamefont{Boukhvalov}},
  \bibinfo{author}{\bibfnamefont{M.~I.} \bibnamefont{Katsnelson}},
  \bibinfo{author}{\bibfnamefont{A.~K.} \bibnamefont{Geim}},
  \bibnamefont{et~al.}, \bibinfo{journal}{Science}
  \textbf{\bibinfo{volume}{323}}, \bibinfo{pages}{610} (\bibinfo{year}{2009}),
  \eprint{http://www.sciencemag.org/content/323/5914/610.full.pdf},
  \urlprefix\url{http://www.sciencemag.org/content/323/5914/610.abstract}.

\bibitem[{\citenamefont{López-Bezanilla
  et~al.}(2009)\citenamefont{López-Bezanilla, Triozon, and
  Roche}}]{Lopez-Bezanilla}
\bibinfo{author}{\bibfnamefont{A.}~\bibnamefont{López-Bezanilla}},
  \bibinfo{author}{\bibfnamefont{F.}~\bibnamefont{Triozon}}, \bibnamefont{and}
  \bibinfo{author}{\bibfnamefont{S.}~\bibnamefont{Roche}},
  \bibinfo{journal}{Nano Letters} \textbf{\bibinfo{volume}{9}},
  \bibinfo{pages}{2537} (\bibinfo{year}{2009}), \bibinfo{note}{pMID: 19505128},
  \eprint{http://pubs.acs.org/doi/pdf/10.1021/nl900561x},
  \urlprefix\url{http://pubs.acs.org/doi/abs/10.1021/nl900561x}.

\bibitem[{\citenamefont{Leconte et~al.}(2011)\citenamefont{Leconte, Soriano,
  Roche, Ordejon, Charlier, and Palacios}}]{nn200558d}
\bibinfo{author}{\bibfnamefont{N.}~\bibnamefont{Leconte}},
  \bibinfo{author}{\bibfnamefont{D.}~\bibnamefont{Soriano}},
  \bibinfo{author}{\bibfnamefont{S.}~\bibnamefont{Roche}},
  \bibinfo{author}{\bibfnamefont{P.}~\bibnamefont{Ordejon}},
  \bibinfo{author}{\bibfnamefont{J.-C.} \bibnamefont{Charlier}},
  \bibnamefont{and} \bibinfo{author}{\bibfnamefont{J.~J.}
  \bibnamefont{Palacios}}, \bibinfo{journal}{ACS Nano}
  \textbf{\bibinfo{volume}{5}}, \bibinfo{pages}{3987} (\bibinfo{year}{2011}),
  \eprint{http://pubs.acs.org/doi/pdf/10.1021/nn200558d},
  \urlprefix\url{http://pubs.acs.org/doi/abs/10.1021/nn200558d}.

\bibitem[{\citenamefont{Rodriguez-Manzo and Banhart}(2009)}]{Rodriquez-Manzo}
\bibinfo{author}{\bibfnamefont{J.~A.} \bibnamefont{Rodriguez-Manzo}}
  \bibnamefont{and} \bibinfo{author}{\bibfnamefont{F.}~\bibnamefont{Banhart}},
  \bibinfo{journal}{Nano Letters} \textbf{\bibinfo{volume}{9}},
  \bibinfo{pages}{2285} (\bibinfo{year}{2009}), \bibinfo{note}{pMID: 19413339},
  \eprint{http://pubs.acs.org/doi/pdf/10.1021/nl900463u},
  \urlprefix\url{http://pubs.acs.org/doi/abs/10.1021/nl900463u}.

\bibitem[{\citenamefont{Lehmann et~al.}(2013)\citenamefont{Lehmann, Ryndyk, and
  Cuniberti}}]{Lehmann}
\bibinfo{author}{\bibfnamefont{T.}~\bibnamefont{Lehmann}},
  \bibinfo{author}{\bibfnamefont{D.~A.} \bibnamefont{Ryndyk}},
  \bibnamefont{and}
  \bibinfo{author}{\bibfnamefont{G.}~\bibnamefont{Cuniberti}},
  \bibinfo{journal}{Phys. Rev. B} \textbf{\bibinfo{volume}{88}},
  \bibinfo{pages}{125420} (\bibinfo{year}{2013}),
  \urlprefix\url{http://link.aps.org/doi/10.1103/PhysRevB.88.125420}.

\bibitem[{\citenamefont{Meyer et~al.}(2008)\citenamefont{Meyer, Kisielowski,
  Erni, Rossell, Crommie, and Zettl}}]{Meyer}
\bibinfo{author}{\bibfnamefont{J.~C.} \bibnamefont{Meyer}},
  \bibinfo{author}{\bibfnamefont{C.}~\bibnamefont{Kisielowski}},
  \bibinfo{author}{\bibfnamefont{R.}~\bibnamefont{Erni}},
  \bibinfo{author}{\bibfnamefont{M.~D.} \bibnamefont{Rossell}},
  \bibinfo{author}{\bibfnamefont{M.~F.} \bibnamefont{Crommie}},
  \bibnamefont{and} \bibinfo{author}{\bibfnamefont{A.}~\bibnamefont{Zettl}},
  \bibinfo{journal}{Nano Letters} \textbf{\bibinfo{volume}{8}},
  \bibinfo{pages}{3582} (\bibinfo{year}{2008}), \bibinfo{note}{pMID: 18563938},
  \eprint{http://pubs.acs.org/doi/pdf/10.1021/nl801386m},
  \urlprefix\url{http://pubs.acs.org/doi/abs/10.1021/nl801386m}.

\bibitem[{\citenamefont{Gomez-Navarro et~al.}(2005)\citenamefont{Gomez-Navarro,
  Pablo, Gomez-Herrero, Biel, Garcia-Vidal, Rubio, and Flores}}]{Gomez-Navarro}
\bibinfo{author}{\bibfnamefont{C.}~\bibnamefont{Gomez-Navarro}},
  \bibinfo{author}{\bibfnamefont{P.~J.~D.} \bibnamefont{Pablo}},
  \bibinfo{author}{\bibfnamefont{J.}~\bibnamefont{Gomez-Herrero}},
  \bibinfo{author}{\bibfnamefont{B.}~\bibnamefont{Biel}},
  \bibinfo{author}{\bibfnamefont{F.~J.} \bibnamefont{Garcia-Vidal}},
  \bibinfo{author}{\bibfnamefont{A.}~\bibnamefont{Rubio}}, \bibnamefont{and}
  \bibinfo{author}{\bibfnamefont{F.}~\bibnamefont{Flores}},
  \bibinfo{journal}{Nat Mater} \textbf{\bibinfo{volume}{4}},
  \bibinfo{pages}{534} (\bibinfo{year}{2005}),
  \urlprefix\url{http://dx.doi.org/10.1038/nmat1414}.

\bibitem[{\citenamefont{Warner et~al.}(2013)\citenamefont{Warner, Liu, He,
  Robertson, and Suenaga}}]{Warmer}
\bibinfo{author}{\bibfnamefont{J.~H.} \bibnamefont{Warner}},
  \bibinfo{author}{\bibfnamefont{Z.}~\bibnamefont{Liu}},
  \bibinfo{author}{\bibfnamefont{K.}~\bibnamefont{He}},
  \bibinfo{author}{\bibfnamefont{A.~W.} \bibnamefont{Robertson}},
  \bibnamefont{and} \bibinfo{author}{\bibfnamefont{K.}~\bibnamefont{Suenaga}},
  \bibinfo{journal}{Nano Letters} \textbf{\bibinfo{volume}{13}},
  \bibinfo{pages}{4820} (\bibinfo{year}{2013}),
  \eprint{http://pubs.acs.org/doi/pdf/10.1021/nl402514c},
  \urlprefix\url{http://pubs.acs.org/doi/abs/10.1021/nl402514c}.

\bibitem[{\citenamefont{Inui et~al.}(1994)\citenamefont{Inui, Trugman, and
  Abrahams}}]{PhysRevB.49.3190}
\bibinfo{author}{\bibfnamefont{M.}~\bibnamefont{Inui}},
  \bibinfo{author}{\bibfnamefont{S.~A.} \bibnamefont{Trugman}},
  \bibnamefont{and} \bibinfo{author}{\bibfnamefont{E.}~\bibnamefont{Abrahams}},
  \bibinfo{journal}{Phys. Rev. B} \textbf{\bibinfo{volume}{49}},
  \bibinfo{pages}{3190} (\bibinfo{year}{1994}).

\bibitem[{\citenamefont{Liang and Sofo}(2012)}]{Liang}
\bibinfo{author}{\bibfnamefont{S.-Z.} \bibnamefont{Liang}} \bibnamefont{and}
  \bibinfo{author}{\bibfnamefont{J.~O.} \bibnamefont{Sofo}},
  \bibinfo{journal}{Phys. Rev. Lett.} \textbf{\bibinfo{volume}{109}},
  \bibinfo{pages}{256601} (\bibinfo{year}{2012}),
  \urlprefix\url{http://link.aps.org/doi/10.1103/PhysRevLett.109.256601}.

\bibitem[{\citenamefont{Palacios and Yndur{\'{a}}in}(2012)}]{Palacios12}
\bibinfo{author}{\bibfnamefont{J.~J.} \bibnamefont{Palacios}} \bibnamefont{and}
  \bibinfo{author}{\bibfnamefont{F.}~\bibnamefont{Yndur{\'{a}}in}},
  \bibinfo{journal}{Physical Review B} \textbf{\bibinfo{volume}{85}},
  \bibinfo{pages}{245443} (\bibinfo{year}{2012}).

\bibitem[{\citenamefont{Duplock et~al.}(2004)\citenamefont{Duplock, Scheffler,
  and Lindan}}]{Duplock}
\bibinfo{author}{\bibfnamefont{E.~J.} \bibnamefont{Duplock}},
  \bibinfo{author}{\bibfnamefont{M.}~\bibnamefont{Scheffler}},
  \bibnamefont{and} \bibinfo{author}{\bibfnamefont{P.~J.~D.}
  \bibnamefont{Lindan}}, \bibinfo{journal}{Phys. Rev. Lett.}
  \textbf{\bibinfo{volume}{92}}, \bibinfo{pages}{225502}
  (\bibinfo{year}{2004}),
  \urlprefix\url{http://link.aps.org/doi/10.1103/PhysRevLett.92.225502}.

\bibitem[{\citenamefont{Soriano et~al.}(2011)\citenamefont{Soriano, Leconte,
  Ordej\'on, Charlier, Palacios, and Roche}}]{Soriano}
\bibinfo{author}{\bibfnamefont{D.}~\bibnamefont{Soriano}},
  \bibinfo{author}{\bibfnamefont{N.}~\bibnamefont{Leconte}},
  \bibinfo{author}{\bibfnamefont{P.}~\bibnamefont{Ordej\'on}},
  \bibinfo{author}{\bibfnamefont{J.-C.} \bibnamefont{Charlier}},
  \bibinfo{author}{\bibfnamefont{J.-J.} \bibnamefont{Palacios}},
  \bibnamefont{and} \bibinfo{author}{\bibfnamefont{S.}~\bibnamefont{Roche}},
  \bibinfo{journal}{Phys. Rev. Lett.} \textbf{\bibinfo{volume}{107}},
  \bibinfo{pages}{016602} (\bibinfo{year}{2011}),
  \urlprefix\url{http://link.aps.org/doi/10.1103/PhysRevLett.107.016602}.

\bibitem[{\citenamefont{Vozmediano et~al.}(2005)\citenamefont{Vozmediano,
  L{\'{o}}pez-Sancho, Stauber, and Guinea}}]{vozmediano:155121}
\bibinfo{author}{\bibfnamefont{M.~A.~H.} \bibnamefont{Vozmediano}},
  \bibinfo{author}{\bibfnamefont{M.~P.} \bibnamefont{L{\'{o}}pez-Sancho}},
  \bibinfo{author}{\bibfnamefont{T.}~\bibnamefont{Stauber}}, \bibnamefont{and}
  \bibinfo{author}{\bibfnamefont{F.}~\bibnamefont{Guinea}},
  \bibinfo{journal}{Physical Review B} \textbf{\bibinfo{volume}{72}},
  \bibinfo{pages}{155121} (\bibinfo{year}{2005}),
  \urlprefix\url{http://link.aps.org/abstract/PRB/v72/e155121}.

\bibitem[{\citenamefont{Son et~al.}(2006{\natexlab{a}})\citenamefont{Son,
  Cohen, and Louie}}]{son:216803}
\bibinfo{author}{\bibfnamefont{Y.-W.} \bibnamefont{Son}},
  \bibinfo{author}{\bibfnamefont{M.~L.} \bibnamefont{Cohen}}, \bibnamefont{and}
  \bibinfo{author}{\bibfnamefont{S.~G.} \bibnamefont{Louie}},
  \bibinfo{journal}{Physical Review Letters} \textbf{\bibinfo{volume}{97}},
  \bibinfo{pages}{216803} (\bibinfo{year}{2006}{\natexlab{a}}),
  \urlprefix\url{http://link.aps.org/abstract/PRL/v97/e216803}.

\bibitem[{\citenamefont{R.Saito et~al.}(1998)\citenamefont{R.Saito,
  G.Dresselhaus, and M.S.Dresselhaus}}]{Saito_book}
\bibinfo{author}{\bibnamefont{R.Saito}},
  \bibinfo{author}{\bibnamefont{G.Dresselhaus}}, \bibnamefont{and}
  \bibinfo{author}{\bibnamefont{M.S.Dresselhaus}},
  \emph{\bibinfo{title}{Physical Properties of Carbon Nanotubes}}
  (\bibinfo{publisher}{Imperial College}, \bibinfo{address}{London},
  \bibinfo{year}{1998}).

\bibitem[{\citenamefont{Hamada et~al.}(1992)\citenamefont{Hamada, Sawada, and
  Oshiyama}}]{Hamada_1992}
\bibinfo{author}{\bibfnamefont{N.}~\bibnamefont{Hamada}},
  \bibinfo{author}{\bibfnamefont{S.-I.} \bibnamefont{Sawada}},
  \bibnamefont{and} \bibinfo{author}{\bibfnamefont{A.}~\bibnamefont{Oshiyama}},
  \bibinfo{journal}{Phys.\ Rev.\ Lett.} \textbf{\bibinfo{volume}{68}},
  \bibinfo{pages}{1579} (\bibinfo{year}{1992}).

\bibitem[{\citenamefont{Dresselhaus}(1996)}]{DresselhausPW_1996}
\bibinfo{author}{\bibfnamefont{M.}~\bibnamefont{Dresselhaus}},
  \bibinfo{journal}{Physics World} \textbf{\bibinfo{volume}{May}},
  \bibinfo{pages}{18} (\bibinfo{year}{1996}).

\bibitem[{\citenamefont{Garc\'{\i}a-Lastra
  et~al.}(2008)\citenamefont{Garc\'{\i}a-Lastra, Thygesen, Strange, and
  Rubio}}]{Thygesen}
\bibinfo{author}{\bibfnamefont{J.~M.} \bibnamefont{Garc\'{\i}a-Lastra}},
  \bibinfo{author}{\bibfnamefont{K.~S.} \bibnamefont{Thygesen}},
  \bibinfo{author}{\bibfnamefont{M.}~\bibnamefont{Strange}}, \bibnamefont{and}
  \bibinfo{author}{\bibfnamefont{A.}~\bibnamefont{Rubio}},
  \bibinfo{journal}{Phys. Rev. Lett.} \textbf{\bibinfo{volume}{101}},
  \bibinfo{pages}{236806} (\bibinfo{year}{2008}),
  \urlprefix\url{http://link.aps.org/doi/10.1103/PhysRevLett.101.236806}.

\bibitem[{\citenamefont{Khalfoun et~al.}(2014)\citenamefont{Khalfoun, Lambin,
  and Henrard}}]{Khalfoun}
\bibinfo{author}{\bibfnamefont{H.}~\bibnamefont{Khalfoun}},
  \bibinfo{author}{\bibfnamefont{P.}~\bibnamefont{Lambin}}, \bibnamefont{and}
  \bibinfo{author}{\bibfnamefont{L.}~\bibnamefont{Henrard}},
  \bibinfo{journal}{Phys. Rev. B} \textbf{\bibinfo{volume}{89}},
  \bibinfo{pages}{045407} (\bibinfo{year}{2014}),
  \urlprefix\url{http://link.aps.org/doi/10.1103/PhysRevB.89.045407}.

\bibitem[{\citenamefont{Son et~al.}(2006{\natexlab{b}})\citenamefont{Son,
  Cohen, and Louie}}]{Louie}
\bibinfo{author}{\bibfnamefont{Y.-W.} \bibnamefont{Son}},
  \bibinfo{author}{\bibfnamefont{M.~L.} \bibnamefont{Cohen}}, \bibnamefont{and}
  \bibinfo{author}{\bibfnamefont{S.~G.} \bibnamefont{Louie}},
  \bibinfo{journal}{Nature} \textbf{\bibinfo{volume}{444}},
  \bibinfo{pages}{347} (\bibinfo{year}{2006}{\natexlab{b}}),
  \urlprefix\url{http://www.nature.com/nature/journal/v444/n7117/full/nature05%
180.html}.

\bibitem[{\citenamefont{Santos et~al.}(2012)\citenamefont{Santos, Ayuela,
  Chico, and Artacho}}]{magnetic_bilayer}
\bibinfo{author}{\bibfnamefont{H.}~\bibnamefont{Santos}},
  \bibinfo{author}{\bibfnamefont{A.}~\bibnamefont{Ayuela}},
  \bibinfo{author}{\bibfnamefont{L.}~\bibnamefont{Chico}}, \bibnamefont{and}
  \bibinfo{author}{\bibfnamefont{E.}~\bibnamefont{Artacho}},
  \bibinfo{journal}{Phys. Rev. B} \textbf{\bibinfo{volume}{85}},
  \bibinfo{pages}{245430} (\bibinfo{year}{2012}),
  \urlprefix\url{http://link.aps.org/doi/10.1103/PhysRevB.85.245430}.

\bibitem[{\citenamefont{Soler et~al.}(2002)\citenamefont{Soler, Artacho, Gale,
  Garc{\'{i}}a, Junquera, Ordej{\'{o}}n, and S{\'{a}}nchez-Portal}}]{SIESTA}
\bibinfo{author}{\bibfnamefont{J.~M.} \bibnamefont{Soler}},
  \bibinfo{author}{\bibfnamefont{E.}~\bibnamefont{Artacho}},
  \bibinfo{author}{\bibfnamefont{J.~D.} \bibnamefont{Gale}},
  \bibinfo{author}{\bibfnamefont{A.}~\bibnamefont{Garc{\'{i}}a}},
  \bibinfo{author}{\bibfnamefont{J.}~\bibnamefont{Junquera}},
  \bibinfo{author}{\bibfnamefont{P.}~\bibnamefont{Ordej{\'{o}}n}},
  \bibnamefont{and}
  \bibinfo{author}{\bibfnamefont{D.}~\bibnamefont{S{\'{a}}nchez-Portal}},
  \bibinfo{journal}{Journal of Physics: Condensed Matter}
  \textbf{\bibinfo{volume}{14}}, \bibinfo{pages}{2745} (\bibinfo{year}{2002}),
  \urlprefix\url{http://stacks.iop.org/0953-8984/14/i=11/a=302}.

\bibitem[{Dat()}]{Datos}
\bibinfo{note}{Spin-polarized calculations were performed using the functional
  revPBE\cite{revPBE}, but we have checked that other functional preserves the
  main features found. Norm conserving Troullier-Martins pseudopotential and a
  $\xi$ double singly polarized basis set have been used for describing the
  core-electrons and the valence electrons, respectively. The real-space grid
  for matrix-element computations uses an energy cutoff of $400$ Ry.}

\bibitem[{\citenamefont{Lieb}(1989)}]{Lieb}
\bibinfo{author}{\bibfnamefont{E.~H.} \bibnamefont{Lieb}},
  \bibinfo{journal}{Phys. Rev. Lett.} \textbf{\bibinfo{volume}{62}},
  \bibinfo{pages}{1201} (\bibinfo{year}{1989}),
  \urlprefix\url{http://link.aps.org/doi/10.1103/PhysRevLett.62.1201}.

\bibitem[{\citenamefont{Chico et~al.}(1996)\citenamefont{Chico, Benedict,
  Louie, and Cohen}}]{Chico3}
\bibinfo{author}{\bibfnamefont{L.}~\bibnamefont{Chico}},
  \bibinfo{author}{\bibfnamefont{L.~X.} \bibnamefont{Benedict}},
  \bibinfo{author}{\bibfnamefont{S.~G.} \bibnamefont{Louie}}, \bibnamefont{and}
  \bibinfo{author}{\bibfnamefont{M.~L.} \bibnamefont{Cohen}},
  \bibinfo{journal}{Phys. Rev. B} \textbf{\bibinfo{volume}{54}},
  \bibinfo{pages}{2600} (\bibinfo{year}{1996}),
  \urlprefix\url{http://link.aps.org/doi/10.1103/PhysRevB.54.2600}.

\bibitem[{\citenamefont{Jacob and Palacios}(2011)}]{Jacob11}
\bibinfo{author}{\bibfnamefont{D.}~\bibnamefont{Jacob}} \bibnamefont{and}
  \bibinfo{author}{\bibfnamefont{J.~J.} \bibnamefont{Palacios}},
  \bibinfo{journal}{J Chem Phys} \textbf{\bibinfo{volume}{134}},
  \bibinfo{pages}{044118} (\bibinfo{year}{2011}).

\bibitem[{\citenamefont{Munoz-Rojas et~al.}(2006)\citenamefont{Munoz-Rojas,
  Jacob, Fernandez-Rossier, and Palacios}}]{Munoz-Rojas06}
\bibinfo{author}{\bibfnamefont{F.}~\bibnamefont{Munoz-Rojas}},
  \bibinfo{author}{\bibfnamefont{D.}~\bibnamefont{Jacob}},
  \bibinfo{author}{\bibfnamefont{J.}~\bibnamefont{Fernandez-Rossier}},
  \bibnamefont{and} \bibinfo{author}{\bibfnamefont{J.~J.}
  \bibnamefont{Palacios}}, \bibinfo{journal}{Physical Review B}
  \textbf{\bibinfo{volume}{74}}, \bibinfo{pages}{195417}
  (\bibinfo{year}{2006}), \eprint{cond-mat/0608720v1}.

\bibitem[{\citenamefont{Santos et~al.}(2009)\citenamefont{Santos, Chico, and
  Brey}}]{Santos}
\bibinfo{author}{\bibfnamefont{H.}~\bibnamefont{Santos}},
  \bibinfo{author}{\bibfnamefont{L.}~\bibnamefont{Chico}}, \bibnamefont{and}
  \bibinfo{author}{\bibfnamefont{L.}~\bibnamefont{Brey}},
  \bibinfo{journal}{Phys. Rev. Lett.} \textbf{\bibinfo{volume}{103}},
  \bibinfo{pages}{086801} (\bibinfo{year}{2009}),
  \urlprefix\url{http://link.aps.org/doi/10.1103/PhysRevLett.103.086801}.

\bibitem[{\citenamefont{Zhang and Yang}(1998)}]{revPBE}
\bibinfo{author}{\bibfnamefont{Y.}~\bibnamefont{Zhang}} \bibnamefont{and}
  \bibinfo{author}{\bibfnamefont{W.}~\bibnamefont{Yang}},
  \bibinfo{journal}{Phys. Rev. Lett.} \textbf{\bibinfo{volume}{80}},
  \bibinfo{pages}{890} (\bibinfo{year}{1998}),
  \urlprefix\url{http://link.aps.org/doi/10.1103/PhysRevLett.80.890}.

\end{thebibliography}

\end{document}